\documentstyle[prl,aps,preprint,epsf]{revtex}
\begin{document}
\newcommand{\dd}{{\mbox{\rm d}}}
\newcommand{\rC}{{\mbox{\rm C}}}
\newcommand{\rR}{{\mbox{\rm R}}}
\newcommand{\rZ}{{\mbox{\rm Z}}}
\newcommand{\cF}{{\cal F}}
\newcommand{\cH}{{\cal H}_-}
\newcommand{\cS}{{\cal S}}
\newcommand{\cT}{{\cal T}}
\setcounter{page}{1}
\draft
\def\footnoterule{\kern-3pt \hrule width\hsize \kern3pt}
\title{Minimum uncertainty for antisymmetric wave functions} 

\author{L.L. Salcedo\footnote{Email address: {\tt salcedo@goliat.ugr.es}}}
\address{Departamento de F\'{\i}sica Moderna \\
Universidad de Granada \\
E-18071 Granada, Spain \\
{~}}

\maketitle

\thispagestyle{empty}

\begin{abstract}

We study how the entropic uncertainty relation for position and
momentum conjugate variables is minimized in the subspace of
one-dimensional antisymmetric wave functions. Based partially on
numerical evidence and partially on analytical results, a conjecture
is presented for the sharp bound and for the minimizers. Conjectures
are also presented for the corresponding sharp Hausdorff-Young inequality.

\end{abstract}


\pacs{PACS }

Let $\psi$ be in a square integrable function in $\rR^n$, to represent
the wave function of a quantum-mechanical particle, and let $\rho$ be
its normalized probability density, to wit,
$\rho(x)=|\psi(x)|^2/||\psi||_2^2$, where $||\psi||_p$ denotes the
$p$-norm $\left(\int|\psi(x)|^p\dd^nx\right)^{1/p}$. The information
entropy of $\psi$ (or $\rho$) is defined as
\begin{equation}
S(\psi)=-\int\log(\rho(x))\rho(x)\dd^nx\,.
\end{equation}
It measures the localization of the state in configuration space. A
high entropy implies a low spatial localization and vice versa.
Likewise, one can consider the wave function in momentum space,
defined by the Fourier transform of $\psi$, that is
\begin{equation}
(\cF\psi)(x)=\int e^{2\pi ixy}\psi(y)\dd^ny\,
\end{equation}
for $\psi$ integrable. (The normalization of $\cF$ corresponds to
using units $2\pi\hbar=1$.) We will often use the notation
$\widetilde\psi$ for the Fourier transform of $\psi$. Again, its
information entropy $S(\widetilde\psi)$ is a measure of its momentum
space localization.

As shown by Hirschman~\cite{Hi57} in one dimension and by
Bia{\l}ynicki-Birula and Mycielski in the $n$-dimensional
case~\cite{Bi75}, the basic uncertainty relations of position and
momentum in quantum mechanics can be derived from the following sharp
bound in $L^2(\rR^n)$:
\begin{equation}
S(\psi)+S(\widetilde\psi)\ge n(1-\log 2)\,.
\label{eq:3}
\end{equation}
Indeed, this inequality puts a bound on the maximum localization in
phase space and, in particular, it can be shown to imply the
uncertainty relations of Heisenberg (Weyl-Heinsenberg
inequality)~\cite{Hi57,Bi75}. As stressed by Deutsch~\cite{De83},
entropic uncertainty relations among observables are a more faithful
expression of the quantum-mechanical uncertainty principle than the
customary generalized Heisenberg relations. (See also
\cite{Pa83,Ma88,Ro95} for further details and applications.)

The equality in (\ref{eq:3}) is reached by any Gaussian function and
moreover these are the unique minimizers~\cite{Li90}. Since the
Gaussian can be taken centered at the origin, the same sharp bound
holds in the subspace of even functions. Less obvious is the value of
the sharp bound in the subspace spanned by the odd functions, i.e.,
$\psi(-x)=-\psi(x)$, as well as the form of the associated minimizing
functions. Such question would arise, for instance, in the case of two
electrons in a triplet spin state since the relative coordinate wave
function must be odd. In this paper we will address this problem in
the one dimensional case, $n=1$. For future reference, we will denote
the functional $S(\psi)+S(\widetilde\psi)$ by $\cS(\psi)$ and the
subspace of the odd functions in $L^2(\rR)$ by $\cH$. Thus, we seek to
find the infimum of $\cS$ in the space $\cH$, and also to establish
the form of the possible minimizers, or, more generally, of the
minimizing sequences.

Quite likely the problem just raised is non trivial if treated in a
fully rigorous mathematical manner. In 1957 it was noted by Hirschman
(in the one dimensional case) that the l.h.s. of (\ref{eq:3}) is non
negative; this result follows from the classical Hausdorff-Young
inequality (see e.g.~\cite{Be75}), he then conjectured that the sharp
bound was attained by Gaussian functions~\cite{Hi57}. It was not until
1975 that Beckner~\cite{Be75}, motivated by preliminary results of
Babenko~\cite{Ba61}, established the necessary sharp version of
Hausdorff-Young inequality from which Hirschman-Beckner inequality
(\ref{eq:3}) immediately follows. On the other hand, the problem of
finding sharp bounds in restricted classes of functions, such as
linear subspaces, seems to have deserved less or not attention at
all. Given the difficulty of the problem, we have adopted here an
exploratory approach in order to gather ``experimental'' information
on the minimizing function, namely, by numerically minimizing the
entropy functional. From the point of view of rigorous mathematical
results, this procedure can only yield upper bounds on the sharp
bound, nevertheless it can provide useful insights and help to make
reasonably founded conjectures on the form of the minimizers.  Such
conjectures are presented below.

Let us briefly describe the numerical procedure used. We have
considered the expansion of the elements of $L^2(\rR)$ in terms of the
orthonormal harmonic oscillator basis $\phi_n(x)=h_n(x)e^{-\pi{x^2}}$,
where $h_n(x)$ are the associated Hermite polynomials. Thus
$\psi(x)=\sum_{n=0}^\infty a_n\phi_n(x)$ in the mean.  In this basis
the Fourier transform takes the simple form
$\widetilde\psi(x)=\sum_{n=0}^\infty i^na_n\phi_n(x)$. The entropy
functional $\cS$ is then transformed into a function of the complex
coefficients $a_n$ and the problem consists in minimizing this
function with respect to $a_{2n+1}$, $n=0,1,2,\dots$, keeping
$a_{2n}=0$ and $\sum_{n=0}^\infty|a_n|^2$ finite. To address this
problem we actually consider the following $N$-dimensional subspace of
$\cH$
\begin{equation}
\psi(x)=\sum_{n=0}^{N-1}a_{2n+1}\phi_{2n+1}(x)\,,
\label{eq:4}
\end{equation}
for $N$ as large as possible, then make use of standard numerical
algorithms to look for the minimum of $\cS(\psi)$ in this space. The
numerical minimization algorithms become less efficient as $N$
increases, thus implying a maximum admissible value for $N$ in
practice. The largest space used was that corresponding to $N=128$,
which, of course, yielded the best (i.e., the lowest) entropy, namely,
$\cS(\psi)=0.61370581$. This number, as well as the minimizing
functions itself, is only very weakly dependent on the minimization
method used (e.g. a steepest descend or a simplex algorithm), the
precise value of $N$ and the initial conditions used. Also, we have
checked that the Gaussian minimum $1-\log 2$ is correctly reproduced
if even as well as odd functions are allowed. It turned out that
imposing the conditions $\psi^*=\psi$ and $\widetilde\psi=+i\psi$ did
not result in an increase of the entropy. Analogous restrictions can
be imposed on the Gaussian minimizer in the subspace of even
functions. The minimizing function (for $N=128$ and the above
mentioned restrictions) is shown in Figure~1.

Motivated by the numerical results, we define the following two
one-parameter families of functions,
\begin{eqnarray}
\Phi_a(x) &=& \sum_{n\in\rZ}(-1)^ne^{-\pi
a^2x^2}e^{-\pi(x-n-\frac{1}{2})^2/a^2}\,, \nonumber \\
\Phi^\prime_a(x) &=& \sum_{n\in\rZ}(-1)^ne^{-\pi a^2(n+\frac{1}{2})^2}
e^{-\pi(x-n-\frac{1}{2})^2/a^2}\,,
\label{eq:5}
\end{eqnarray}
where the parameter $a$ takes positive values. Note that $\Phi_a$ and
$\Phi_a^\prime$ are two unrelated functions; the symbol ${}^\prime$ is
used to distinguish them and it does not denote a derivative. Our
preliminary ansatz is that the small $a$ limit of $\Phi_a$ (or
equivalently of $\Phi_a^\prime$) corresponds to a minimizer of $\cS$
in $\cH$. Under this assumption, the numerical curve shown in Figure~1
would be a regularized approximation to the small $a$ limit of
$\Phi_a$. In fact, the numerical curve coincides almost perfectly with
$\Phi_a$ or $\Phi^\prime_a$ for $a=0.29$. The parameter $a$ plays the
role of a regulator in eqs.~(\ref{eq:5}), similar to value of $N$ in
eq.~(\ref{eq:4}).

For convenience we will refer to $(\Phi_a)$ and $(\Phi_a^\prime)$ as
sequences since it is always possible to choose a positive sequence
$(a_n)$ with $\lim_{n\to\infty}a_n =0$ so that $(\Phi_{a_n})$ is a
sequence in the usual sense. Strictly speaking the limits as $a\to 0$
of the $(\Phi_a)$ or $(\Phi^\prime_a)$ do not take place within
$L^2(\rR)$, i.e. in norm. Indeed, their point-wise limit is 0 except
at the points $x_n=n+\frac{1}{2}$, where they take the value 1,
whereas their norms $(||\Phi_a||_2)$ and $(||\Phi_a^\prime||_2)$
converge to $1/\sqrt{2}$, as will be shown below. On the other hand
the limit of $(||\Phi_a-\Phi_a^\prime||_2)$ is 0, thus both sequences
$(\Phi_a)$ and $(\Phi^\prime_a)$ become equivalent for small $a$.

We will introduce the following notation. Let $V$ be a normed vector
space, and let $(x_a)$ and $(y_a)$ be two sequences in $V$ (in the
sense $a\to 0$ and $a$ taking positive values). We will say that they
strongly approach each other if $\lim_{a\to 0}||x_a-y_a||=0$, and this
will be denoted by $x\equiv y$ or $x_a\equiv y_a$. Let us remark that
the sequences are not assumed to be Cauchy sequences, hence nothing is
implied for the limits of $||x_{a_1}-x_{a_2}||$ or
$||x_{a_1}-y_{a_2}||$ as $a_1$ and $a_2$ independently approach
$0$. From the triangle inequality it follows that this is an
equivalence relation. Furthermore, if $x_a\equiv y_a$, it follows that
$\lim_{a\to 0}(||x_a||-||y_a||)=0$, since $|(||x_a||-||y_a||)|\le
||x_a-y_a||$. With this notation $\Phi_a\equiv\Phi_a^\prime$ in
$L^2(\rR)$. This is proved in Lemma 1 below.

The word ``limit'' applied to the sequences $(\Phi_a)$ and
$(\Phi^\prime_a)$ is used here only in an improper sense. The strict
statement, if correct, would be that $(\Phi_a)$ is a minimizing
sequence in $\cH$, that is, one that approaches the infimum of $\cS$
in $\cH$. A calculation, to be discussed in more detail later, shows
that $\lim_{a\to 0}\cS(\Phi_a)= 2(1-\log 2)$, therefore we formulate
the following conjecture:

{\bf Conjecture 1.} {\it The infimum of the functional $\cS$ in $\cH$ is
$2(1-\log 2)$.}

Our best numerical value for $\cS$ ~($0.61370581$) is only slightly
above $2(1-\log 2)$ ($0.61370564$). Let us make some remarks on the
form of the assumed minimizing sequence and its improper limit
$\Phi_0$. Both $\Phi_a$ and $\Phi_a^\prime$ are odd and real functions
and moreover $\cF\Phi_a=i\Phi^\prime_a$ and vice versa, thus
$\cF\Phi_a \equiv i\Phi_a$ in $L^2(\rR)$ (and also
point-wise). $\Phi_0$ is formed by a set of localized states arranged
antisymmetrically around 0 and distributed equidistantly through the
real line. The small scale structures (the so called ultraviolet
region in physics) are narrow Gaussian functions, namely,
$e^{-\pi(x-x_n)^2/a^2}$. Likewise, the large scale structure (infrared
region) is a wide Gaussian function centered at the origin, i.e,
$e^{-\pi a^2x^2}$. We will refer to this overall arrangement as a
singular ``bi-Gaussian'' function. As it will be shown below, the
double Gaussian structure of the minimizer $\Phi_0$ is directly
responsible for the fact that $2(1-\log 2)$ is twice the infimum of
$\cS$ in $L^2(\rR)$, which is saturated by a (simple) Gaussian
function. As we will recall in a moment, minimizing $\cS$ is
equivalent to maximize the Fourier transform operator. For any linear
operator from $L^p(\rR^n)$ into $L^q(\rR^n)$ with a Gaussian kernel,
Lieb has shown that the (unrestricted) maximizer, if any, must be a
Gaussian function~\cite{Li90}. The subspace $\cH$ does not contain
such functions, yet it seems that nevertheless a kind of Gaussian, to
wit, a bi-Gaussian function, is the maximizer also in this case. It is
remarkable how even the seemingly simple constraint $\psi\in\cH$
yields a quite unexpected result, which however is full of structure.

The sharp bound $n(1-\log 2)$ on the entropy $\cS$ follows from
computing the norm of the Fourier transform operator $\cF$ considered
as a linear operator from the space $L^p(\rR^n)$ into its dual
$L^q(\rR^n)$, with $p^{-1}+q^{-1}=1$ and $1<p\le 2 \le
q$~\cite{Hi57}. Indeed, we can define a new functional as
\begin{equation}
\cS_q(\psi)=-\log\left(\frac{||\cF\psi||_q}{||\psi||_p}\right)\,.
\end{equation}
$\cS_q$ vanishes at $q=2$, since $\cF$ is unitary in $L^2(\rR^n)$. The
functional $\cS_q$ is related to the $\cS$ by
\begin{equation}
\cS(\psi)= 4\left.\frac{\dd\cS_q(\psi)}{\dd q}\right|_{q=2}\,,
\label{eq:8}
\end{equation}
where the derivative is a right derivative. Following Hirschman's
argument, let $K_q(V)$ denote the norm of the operator $\cF$
restricted to a subspace $V$ of $L^p(\rR^n)$, i.e.
\begin{equation}
\inf\{\cS_q(\psi), \ \psi\in V\} = -\log K_q(V)\,.
\end{equation}
From eq.~(\ref{eq:8}) and using $K_2(V)=1$, it follows
\begin{equation}
\inf\{\cS(\psi), \ \psi\in V\} = 
-4\left.\frac{\dd K_q(V)}{\dd q}\right|_{q=2}\,.
\label{eq:10}
\end{equation}
As first proved by Beckner~\cite{Be75}, the infimum in $L^p(\rR^n)$ is
reached by Gaussian functions and thus $K_q(L^p(\rR^n))=
(p^{1/p}q^{-1/q})^{n/2}$. In view of Conjecture 1, it is natural to
make the stronger assumption

{\bf Conjecture 2.} {\it The norm of the linear operator $\cF$ from the
space $\cH^p$ of odd functions of $L^p(\rR)$ into $L^q(\rR)$, with $1
< p\le 2$, is $K_q(\cH^p)=p^{1/p}q^{-1/q}$. Correspondingly, the
infimum of $\cS_q$ in $\cH^p$ is $\frac{1}{q}\log q -\frac{1}{p}\log
p$.}

Conjecture 1 follows from this one. A calculation to be detailed below
shows that the sequences $(\Phi_a)$ and $(\Phi^\prime_a)$ yield the
value $\cS_q=\frac{1}{q}\log q -\frac{1}{p}\log p$ as $a$ goes to 0,
thus, according to this conjecture, they are minimizing sequences also
for $\cS_q$ in $\cH^p$.

Conjectures 1 and 2 settle the point (or, more properly, open the
question) of the infimum of $\cS$ and $\cS_q$ in $\cH^p$. As noted, we
do not expect a strict minimizer of $\cS_q$ to exist and we have
instead to consider minimizing sequences, i.e. such that
$\lim_{n\to\infty}\cS_q(\psi_n)=\inf\{\cS_q, \hbox{in}\ \cH^p\}$. To
address this point and the related problem of uniqueness, and also to
give further support to the conjectures, we will now turn to a more
detailed study of the bi-Gaussian ansatzs $\Phi_a$ and $\Phi_a^\prime$
and their admissible generalizations.

Let $D_0$ denote the class of distributions, $d_0(x)$ of the form
\begin{equation}
d_0(x)=\sum_{n\in\rZ}b_n\delta(x-x_n)\,,\quad x_n=x_0+nr\,,\quad
|b_n|=b\,,\quad r,b>0\,,
\end{equation}
for some $x_0$, $r$ and $b$. Here $\delta(x)$ is Dirac's delta
function. Further, let $\psi_2$ be a function in $L^p(\rR^2)$. Then,
for each $a>0$ we will associate to $\psi_2$ two functions $\psi_1$
and $\psi_1^\prime$ in $L^p(\rR)$ by means of the relations
\begin{eqnarray}
\psi_1(x) &=& \int d_0(y)\psi_2\left(ax,\frac{x-y}{a}\right)\dd y
=\sum_{n\in\rZ}b_n\psi_2\left(ax,\frac{x-x_n}{a}\right) \,, \\
\psi^\prime_1(x) &=& \int d_0(y)\psi_2\left(ay,\frac{x-y}{a}\right)\dd
y =\sum_{n\in\rZ}b_n\psi_2\left(ax_n,\frac{x-x_n}{a}\right) \,.
\end{eqnarray}
We will use the notations $\langle d_0,\psi_2\rangle_a$ and $\langle
d_0,\psi_2\rangle^\prime_a$ to the denote the defining constructions
of $\psi_1$ and $\psi_1^\prime$ respectively. Rather than state the
more general conditions on $\psi_2$ for the above definitions to make
sense, we will restrict $\psi_2$ to the Schwartz space of fast
decreasing $C^\infty(\rR^2)$ functions, on which the tempered
distributions are defined. This space will be denoted by $\cT$, and
will be considered as a subspace of $L^p(\rR^2)$. It has several
useful properties: it is dense in $L^p(\rR^2)$, is invariant under
Fourier transform, and their elements are sufficiently regular for our
purposes, in particular, the defining series of $\psi_1$ and
$\psi_1^\prime$ exist and are absolutely and uniformly convergent for
given $a$.

Both definitions are related by using either $ax$ or $ax_n$ as the
first argument of $\psi_2$. We will be interested throughout in the
limit of small positive $a$. In this limit, and for each $n$, the
terms $\psi_2(ax,(x-x_n)/a)$ and $\psi_2(ax_n,(x-x_n)/a)$ vanish
unless $x-x_n$ is of order $a$, thus both definitions become
equivalent. More precisely, they strongly approach each other as $a$
goes to 0, i.e. 
\begin{equation}
\langle d_0,\psi_2\rangle_a \equiv\langle d_0,\psi_2\rangle_a^\prime
\quad\mbox{in\ } L^p(\rR)\,.
\end{equation}
This statement is meaningful since the normalizations of $\psi_1$ and
$\psi_1^\prime$ are well-defined; they have finite $p$-norm as $a$
goes to 0. This is proved in Lemma 1 and Proposition 1 below.

Perhaps the best way of understanding the constructions $\psi_1$ and
$\psi_1^\prime$ is to consider the case of a separable function
$\psi_2(x,y)=\alpha(x)\beta(y)$. The definition of $\psi_1$
corresponds to make a convolution of $d_0$ with $\beta(x/a)$ and then
multiply by $\alpha(ax)$, whereas in $\psi_1^\prime$ the
multiplication is performed in the first place and the convolution is
done next. In the limit of small $a$ both operations commute, i.e.,
$\alpha(\beta*d_0)\equiv\beta*(\alpha d_0)$. The function $\alpha$
describes the large scale profile of the function, whereas $\beta$
gives the small scale structure of $\psi_1$. For arbitrary functions
$\psi_2$, which can always be considered as a linear combination of
separable ones, those roles are played by $\psi_2(x,0)$ and
$\psi_2(0,x)$, respectively.

The bi-Gaussian ansatzs $\Phi_a$ and $\Phi_a^\prime$ are of the form
$\langle d_0,\psi_2\rangle_a$ and $\langle d_0,\psi_2\rangle_a^\prime$
with
\begin{eqnarray}
d_0(x) &=& \sum_{n\in\rZ}(-1)^n\delta(x-n-\frac{1}{2})\,, \label{eq:13}
\\ \psi_2(x,y) &=& \exp(-\pi(x^2+y^2))\,,\label{eq:14}
\end{eqnarray}
i.e. $b_n=(-1)^n$ and $x_n=n+\frac{1}{2}$.

The small $a$ limit of $\psi_1$ or $\psi_1^\prime$ does not take place
in $L^p(\rR)$, nevertheless, after an appropriate renormalization,
there is a weak limit as a distribution, namely
\begin{equation}
\lim_{a\to 0} \frac{1}{a}\psi_1(x)= \lim_{a\to 0}
\frac{1}{a}\psi_1^\prime(x)= Kd_0(x)\,,
\end{equation}
where the constant $K=\int\psi_2(0,x)\dd x$. This follows from
considering the integrals $\int\psi_1(x)f(x)\dd x$ or
$\int\psi_1^\prime(x)f(x)\dd x$ for an arbitrary test function $f$,
after the change of variables $x\to ax+y$.

Let us compute the $p$-norm of $\psi_1$ in the limit $a\to 0$. Since
in this limit the overlap among different terms of the defining series
of $\psi_1$ is negligible, for each value of $x$ at most one term of
the series is relevant. This can be formulated as follows. For given
$d_0$ in $D_0$ and each integer $n$, let $I_n$ be interval
$[x_n-\frac{1}{2}r,x_n+\frac{1}{2}r)$ and let $\varphi_n(x)$ denote
the characteristic function of $I_n$. Then
\begin{equation}
\widehat{\psi}_1(x)=
\sum_{n\in\rZ}b_n\psi_2\left(ax,\frac{x-x_n}{a}\right)\varphi_n(x)
\,,\quad \widehat{\psi}_1^\prime(x)=
\sum_{n\in\rZ}b_n\psi_2\left(ax_n,\frac{x-x_n}{a}\right)\varphi_n(x)
\end{equation}
represent the truncated functions, obtained keeping only the relevant
$n$ for each $x$, i.e., such that $x\in I_n$.

The functions $\psi_1$, $\psi_1^\prime$ and their truncated versions
are all equivalent:

{\bf Lemma 1.} {\it Let $\psi_2\in\cT$, $d_0\in D_0$ and
$p\ge1$, then
$\psi_1\equiv\psi_1^\prime\equiv\widehat\psi_1\equiv\widehat\psi_1^\prime$
in $L^p(\rR)$.}

The proof is given in the Appendix.

We can now compute the $p$-norms of $\psi_1$ in the limit of small $a$.

{\bf Proposition 1.} {\it Let $\psi_2\in\cT$, $d_0\in D_0$ and
$p\ge 1$, then}
\begin{equation}
\lim_{a\to 0}||\psi_1||_p =\lim_{a\to 0}||\psi_1^\prime||_p
=\frac{b}{r^{1/p}}\,||\psi_2||_p \,.
\label{eq:19}
\end{equation}

{\it Proof.} Due to the previous lemma, it is enough to compute $\lim_{a\to
0}||\widehat\psi_1^\prime||_p$.
\begin{eqnarray}
||\widehat{\psi}_1^\prime||_p^p &=&
b^p\sum_{n\in\rZ}\int_{I_n}\left|\psi_2\left(ax_n,(x-x_n)/a\right)\right|^p\dd
x \nonumber\\ 
&=& b^p a\sum_{n\in\rZ}
\int_{-r/2a}^{r/2a}\left|\psi_2\left(ax_n,x\right)\right|^p\dd x\,.
\end{eqnarray}
Due to Lemma 1, the limits of the integral can be extended to
$\pm\infty$. Next, we can use that for a Riemann integrable function
$f$, $\int f(x)\dd x= \lim_{h\to 0} \sum_{n\in\rZ}h f(x_0+h n)$. Thus,
\begin{equation}\lim_{a\to 0}||\widehat{\psi}_1^\prime||_p^p =
\frac{b^p}{r}\int\left|\psi_2(y,x)\right|^p\dd x\dd y \,.
\end{equation}
This proves the proposition.

{\bf Corollary 1.} {\it For $\psi_2,\phi_2\in\cT$, $d_0\in D_0$ and
$p\ge 1$, $\langle d_0,\psi_2\rangle_a \equiv \langle
d_0,\phi_2\rangle_a$ in $L^p(\rR)$ if and only if $\psi_2 =\phi_2$.}

A straightforward calculation shows that the Fourier transforms of
$\psi_1$ and $\psi_1^\prime$ are related to that of $\psi_2$ as
\begin{equation}
\cF\langle d_0,\psi_2\rangle_a=
\langle\widetilde{d}_0,T\widetilde{\psi}_2\rangle_a^\prime\,, \qquad
\cF\langle d_0,\psi_2\rangle_a^\prime=
\langle\widetilde{d}_0,T\widetilde{\psi}_2\rangle_a\,,
\end{equation}
where $T$ denotes the transposition operator,
$T\psi_2(x,y)=\psi_2(y,x)$.

In order to proceed, we will consider admissible only the
distributions $d_0$ in $D_0$ whose Fourier transform
$\widetilde{d}_0(x)$ is again in the class $D_0$, that is
\begin{equation}
\widetilde{d}_0(x)=\sum_{n\in\rZ}\widetilde{b}_n\delta(x-\widetilde{x}_n)\,,
\quad\widetilde{x}_n= \widetilde{x}_0+n\widetilde{r}\,,\quad
|\widetilde{b}_n|=\widetilde{b}\,,
\end{equation}
for some $\widetilde{x}_0$, $\widetilde{r}$ and $\widetilde{b}$. The
admissible distributions span the space $D_0^*:= D_0\cap\cF^{-1}D_0$.
Then, recalling that $\cF$ is a bijection in $\cT$, Lemma 1 and
Proposition 1 apply to $\widetilde{\psi}_1$ and
$\widetilde{\psi}^\prime_1$. This immediately leads to

{\bf Proposition 2.} {\it For $\psi_2\in\cT$, $d_0\in D_0^*$ and
$1\le p\le 2$, $p^{-1}+q^{-1}=1$,}
\begin{eqnarray}
\lim_{a\to 0}\cS_q(\psi_1) &=& \cS_q(\psi_2)+C_q\,,\quad C_q =
\log\left(\frac{b}{\widetilde{b}}
\frac{\widetilde{r}^{1/q}}{r^{1/p}}\right)\,,
\\ \lim_{a\to 0}\cS(\psi_1) &=& \cS(\psi_2)+C\,,\quad
C=-\log\left(r\widetilde{r}\right)\,.
\end{eqnarray}
In both cases, the first term depends only on $\psi_2$ and the second
one only on $d_0$. Furthermore, if $\psi_2(x,y)$ happens to be
separable as $\alpha(x)\beta(y)$, the entropies also split as the sum
of the entropies corresponding to the infrared part $\alpha$ plus the
ultraviolet part $\beta$. Let us denote by ${\cal H}(d_0)$ the
(improper) subspace of $L^2(\rR)$ spanned by the functions $\psi_1$,
for given $d_0$, in the limit of small $a$. From these formulae
follows that the minimum entropy in ${\cal H}(d_0)$ corresponds to
those $\psi_1$ associated to a Gaussian $\psi_2$, i.e. $\psi_1$ must
be a bi-Gaussian function. Therefore, 

{\bf Corollary 2.} {\it Under the same assumptions as in Proposition
2, the following bounds are sharp in ${\cal H}(d_0)$ and are attained by
$\psi_2$ Gaussian.}
\begin{eqnarray}
\lim_{a\to 0}\cS_q(\psi_1) &\ge& \frac{1}{q}\log q -\frac{1}{p}\log p
+ C_q\,,\label{eq:25}\\
\lim_{a\to 0}\cS(\psi_1) &\ge& 2(1-\log 2) + C\,.
\label{eq:26}\end{eqnarray}

On the other hand, $\cS_q(\psi_1)$ is bounded from below by its
infimum in $L^p(\rR)$, namely, $\frac{1}{2q}\log q -\frac{1}{2p}\log
p$, hence it follows that any $d_0$ in $D_0^*$ must satisfy the
following inequalities
\begin{eqnarray}C_q &\ge& -\frac{1}{2q}\log q +\frac{1}{2p}\log p\quad
(1<p\le 2)\,, \nonumber \\
C &\ge& -1 +\log 2\,.
\label{eq:30}
\end{eqnarray}

The distribution $d_0$ in eq.~(\ref{eq:13}) satisfies $\cF d_0=id_0$.
This is easily proved using Poisson's summation formula
$\sum_{n\in\rZ}\exp(i2\pi nx)= \sum_{n\in\rZ}\delta(x+n)$. Thus this
$d_0$ is admissible and the sharp bounds apply with
$r=\widetilde{r}=b=\widetilde{b}=1$, i.e., with $C_q=C=0$. This is
consistent with Conjecture 2 since $\Phi_a$ is a bi-Gaussian odd
function. Moreover, $\widetilde{\Phi}_a = i\Phi_a^\prime$ and
$\widetilde{\Phi}^\prime_a = i\Phi_a$. This follows from
$\widetilde\psi_2(x,y)=\psi_2(y,x)$ for $\psi_2$ in eq.~(\ref{eq:14}).

As a further check of Conjecture 2, let us show that, for the
admissible $d_0$, the functional $\cS_q$ is stationary at $\psi_1$,
when $\psi_1$ is a singular bi-Gaussian function.

{\bf Lemma 2.} {\it Let $d_0\in D_0^*$, then
$b^2/r=\widetilde{b}^2/\widetilde{r}$.}

{\it Proof.} This follows from using that $\cF$ is unitary in $L^2(\rR^n)$,
thus, for any $\psi_2\in\cT$,
\begin{equation}
\frac{b^2}{r}||\psi_2||_2^2=\lim_{a\to 0}||\psi_1||_2^2=
\lim_{a\to 0}||\widetilde\psi_1||_2^2
=\frac{\widetilde{b}^2}{\widetilde{r}}||T\widetilde{\psi}_2||_2^2
=\frac{\widetilde{b}^2}{\widetilde{r}}||\psi_2||_2^2\,.
\end{equation}

{\bf Proposition 3.} {\it Let $\psi_2$ be a Gaussian function, and
$d_0\in D_0^*$, then $\cS$ and $\cS_q$ are stationary at
$\psi_1=\langle d_0,\psi_2\rangle_a$ in the limit $a\to 0$.}

{\it Proof.} Let us consider a first order variation of $\psi$. The first
order variation of $\cS_q$ in $L^p(\rR^n)$ is easily computed from its
definition, yielding
\begin{equation}
\delta\cS_q(\psi)=
\mbox{Re}\int\delta\psi^*(x)\left(G_p-\cF^{-1}G_q\cF\right)
\psi(x)\dd^nx\,,
\end{equation}
where we have introduced the non linear operator $G_s$
\begin{equation}
G_s\psi(x)= \frac{\psi(x)|\psi(x)|^{s-2}}{||\psi||_s^s}\,.
\end{equation}
Since any Gaussian function is a minimizer of $\cS_q$, it follows that
$(G_p-\cF^{-1}G_q\cF)\psi_2$ vanishes identically when $\psi_2$ is
Gaussian. On the other hand, since $\psi_1$ bi-Gaussian is a minimizer
in the subspace ${\cal H}(d_0)$, $\delta\cS_q(\psi_1)$ will also
vanish if the variation $\delta\psi^*$ is in this subspace. What has
to be proved is that in fact $\delta\cS_q(\psi_1)$ vanishes under
arbitrary variations, in the limit of small $a$. This is equivalent to
prove that $(G_p-\cF^{-1}G_q\cF)\psi_1$ vanishes when $\psi_1$ is a
singular bi-Gaussian function. From arguments entirely similar to
those used to establish Proposition 1, it follows
\begin{equation}
G_s\langle d_0,\psi_2\rangle_a \equiv \frac{r}{b^2}\left\langle d_0,
G_s{\psi}_2\right\rangle_a\,,
\end{equation}
from where it is finally obtained
\begin{equation}
\left(G_p-\cF^{-1}G_q\cF\right)
\left\langle d_0,\psi_2\right\rangle_a
\equiv \left\langle d_0,
\left(\frac{r}{b^2}G_p
-\frac{\widetilde{r}}{\widetilde{b}^2}\cF^{-1}G_q\cF\right)\psi_2
\right\rangle_a\,.
\end{equation}
Now, from Lemma 2, $r/b^2$ equals $\widetilde{r}/\widetilde{b}^2$; this
quantity factors out and the right-hand side vanishes for $\psi_2$
Gaussian. This completes the proof.

Let us note that the inequalities~(\ref{eq:30}), as well as Lemma 2
are statements on the space $D_0^*$ only, independent of the
construction $\langle d_0,\psi_2\rangle_a$. This construction,
however, defines a regularization of $d_0$ which has proven useful to
establish properties in $D_0^*$.

It is also interesting to note that a similar construction to that of
$\Phi_a$ can be carried out for the space of even functions, using the
same $\psi_2$ given in eq.~(\ref{eq:14}) and the distribution
\begin{equation}
d_0(x) =\sum_{n\in\rZ}\delta(x-n)\,, \label{eq:28}
\end{equation}
which satisfies $\widetilde{d}_0=d_0$. All the previous arguments
apply here and the same sharp bound for $\cS_q$ in ${\cal H}(d_0)$ is
obtained as for the case of $d_0$ odd. Of course, the corresponding
bi-Gaussian is known not to be a minimizer of the even functions
subspace and at most it can be a relative minimum.

An immediate question is that of the uniqueness of the minimizing
sequence. To study this point we have first to consider the symmetries
of $\cS$ and $\cS_q$. In $L^2(\rR^n)$, the functional $\cS$ is
invariant under: ({\it i}) multiplication by a non-vanishing complex
constant, $\psi(x)\mapsto \lambda\psi(x)$, ({\it ii}) affine regular
transformations, $\psi(x)\mapsto\psi(Ax+b)$, ({\it iii}) complex
conjugation $\psi(x)\mapsto\psi^*(x)$, and ({\it iv}) Fourier
transform, $\psi(x)\mapsto\widetilde\psi(x)$. The functional $\cS_q$
in $L^p(\rR^n)$ is invariant under the transformations {\it (i)-(iii)}
above, whereas under Fourier transform it satisfies $\cS_p(\cF\psi)=
-\cS_q(\psi)$, provided the corresponding norms exist. In $\cH^p$
translation invariance does not exist and linear transformations
consist only of dilatations. The improper minimizer $\Phi_0$ is
invariant under complex conjugation and Fourier transform but breaks
dilatation and normalization invariances (as also does any non trivial
function in $L^2(\rR^n)$). In physics language, these two symmetries
are spontaneously broken. A similar statement can be made for the
Gaussian minimizers in $L^p(\rR^n)$. Let us remark, however, that the
group of symmetries generated by transformations {\it (i)-(iv)} does
not act transitively on the set of minimizers of $\cS$ in $L^2(\rR)$,
since complex (rather than real) affine transformations would needed
to connect two arbitrary Gaussian functions. Likewise, the previous
symmetries applied to $\Phi_0$ do not exhaust the set of minimizers,
and in fact the symmetry group in ${\cal H}(d_0)$ is even larger;
e.g. two independent dilatations applied to $d_0$ and $\psi_2$ still
define an symmetry transformation which acts effectively on ${\cal
H}(d_0)$ (always meaning in the limit of small $a$ for which ${\cal
H}(d_0)$ has been defined).

In passing, note that under a dilatation $d_0(x)\to\mu^{1/2}d_0(\mu x)$,
($\mu$ positive), the quantities $b$, $\widetilde{b}$, $r$ and
$\widetilde{r}$ scale as $\mu^{-1/2}b$, $\mu^{1/2}\widetilde{b}$,
$\mu^{-1}r$ and $\mu\widetilde{r}$, respectively. Thus the
quantities $r/b^2$, $C_q$ and $C$ are dilatation invariant, as they should.

As noted above, the sequence $(\Phi_a)$ is not convergent in
$L^p(\rR)$ and it cannot be made convergent by a suitable
($a$-dependent) renormalization of $\Phi_a$ since its weak limit is
the singular distribution $d_0$. Thus it is not a Cauchy sequence; two
elements $\Phi_{a_1}$ and $\Phi_{a_2}$ need not be near each other in
the strong topology, even for arbitrarily small values of $a_1$ and
$a_2$. That is, the sequence does not even approach itself in the mean
and hence a precise definition is needed to state that some other
minimizing sequence must approach this one. Besides, note that given a
minimizing sequence, one can apply independent arbitrary symmetry
transformations for each value of $a$ and still have a minimizing
sequence.  This implies that a minimizing sequence needs not approach
strongly $(\Phi_a)$ or more generally $(\langle d_0,\psi_2\rangle_a)$
for fixed (i.e. $a$-independent) $d_0$ and $\psi_2$.

The numerical calculation shows that the infimum in $\cH$ can be
achieved in the subspace $\cF=+i$, whereas that corresponding to the
subspace $\cF=-i$ is larger. This suggests that Fourier transform
invariance is not spontaneously broken, that is, that after an
appropriate dilatation, the minimizer can be brought to the space
$\cF=+i$ (note that $\cF$ is not invariant under dilatations). This is
similar to the problem of minimizing $\cS$ in $L^2(\rR)$; a minimizer
(a Gaussian) is not necessarily an even function, but it can be
brought to one after a suitable translation. In the case of $\cH^p$,
not existing a true minimizer, it is important to specify in which
sense the minimizer must satisfy the condition $\cF\psi=+i\psi$
(assuming our conjecture of unbroken Fourier transform invariance to
hold). One can expect that the condition is satisfied in the weak
sense. This is consistent with the fact that $a^{-1}\psi_1$ weakly
converges to $d_0$. It cannot be expected, however, to hold in strong
sense for an arbitrary minimizing sequence. This can be seen noting
that every centered Gaussian $\psi_2$, together with $d_0$ in
eq.~(\ref{eq:13}), would yield a minimizing sequence in $\cH^p$. By a
centered Gaussian, it is meant a function of the form $\psi_2(x,y)=
N\exp(-\frac{1}{2}Ax^2- \frac{1}{2}By^2 -Cxy)$, where $N$, $A$, $B$
and $C$ are complex numbers, $N$ is non vanishing and the real part of
$\frac{1}{2}Ax^2+\frac{1}{2}By^2 +Cxy$ is a positive definite
quadratic form. In this case $\psi_1=\langle d_0,\psi_2\rangle_a$ is
an antisymmetric bi-Gaussian function. Then,
$\widetilde{\psi}_1^\prime-i\psi_1 = i\langle
d_0,T\widetilde{\psi}_2-\psi_2\rangle_a$, whose norm (from Corollary
1) does not go to 0 unless $\widetilde\psi_2=T\psi_2$, and this
equality does not hold for an arbitrary centered Gaussian $\psi_2$.

Another consideration follows from noting that the information entropy
$S$ of $d_0$ is undefined; each single delta function has entropy
minus infinity since they correspond to a maximal localization,
however, the fact that this localization can occur in any of the
points $x_n$ with equal probability adds a plus infinity to the
entropy yielding a undefined value. It follows that the value of $\cS$
or $\cS_q$ for a sequence in ${\cal H}(d_0)$ depends not only on its
weak limit, $d_0$, but also on the particular shape of the functions:
the true minimizer must be Gaussian-like.

Perhaps it will be useful to illustrate the situation with an
example. Consider the minimization of the functional
$F(\psi)=F_0(\psi)+F_1(\psi)$ on $L^2(\rR)$, where
\begin{equation}
F_0(\psi)= S(\psi)-\frac{1}{2}\log\langle (x-\langle
x\rangle_\psi)^2\rangle_\psi\,,\quad F_1(\psi)= \langle (x-\langle
x\rangle_\psi)^2\rangle_\psi\,,
\end{equation}
and $\langle f(x)\rangle_\psi$ means $\int f(x)\rho(x)\dd x$. $F_0$
has been adjusted so that it is invariant under dilatations, whereas,
$F_1$ is minimized by functions as narrow as possible. Therefore, we
can proceed by classifying the space of functions by their value of
$F_1$, and choose the minimizer of $F_0$ in each class. A simple
calculation, using Lagrange multipliers, shows that the minimizer is a
Gaussian located anywhere and with arbitrary normalization and a
well-defined width $a$. This gives $F_1= a^2/2\pi$ and
$F_0=\frac{1}{2}+\frac{1}{2}\log 2\pi$. Next, in order to minimize
$F_1$, we should take $a\to 0$. The infimum of $F$ is then
$\frac{1}{2}+\frac{1}{2}\log 2\pi$. The absolute minimizer does not
exist in $L^2(\rR)$, but a minimizing sequence must approach in some
sense the sequence $\psi_a(x)=\exp(-\pi x^2/2a^2)$ in the limit $a\to
0$, modulo normalization and location. Furthermore, the corresponding
probability density $\rho_a(x)$ must approach the distribution
$\delta(x)$, again modulo translations.

After these considerations, we will state our conjecture on the form
of the minimizer of $\cS_q$ in $\cH^p$. Essentially, it is that a
minimizer must necessarily be a singular bi-Gaussian in the space
$\cF=+i$ (in the weak sense) and modulo dilatations. To put this
conjecture in precise terms, let $d_0$ denote precisely the
distribution in eq.~(\ref{eq:13}) and let $\psi^\mu(x)$ denote
$\psi(\mu x)$, where $\mu>0$ and $\psi\in L^p(\rR)$.

{\bf Conjecture 3.} {\it Let $(\psi_a)$ be a minimizing sequence for
$\cS_q$ in $\cH^p$ (in the sense of $a\to 0$ and the parameter $a$
taking positive values). Then, (a) there is a sequence of positive
numbers $(\mu_a)$ and a sequence of complex numbers $(\lambda_a)$ such
that $(\lambda_a\psi_a^{\mu_a})$ converges weakly to $d_0$. According
to (a), let us assume, without loss of generality, that $(\psi_a)$ has
this property with $\mu_a=1$ and that it is normalized to unity,
$||\psi_a||_p=1$. Then,(b) there is a sequence of centered Gaussian
functions $(\psi_{2,a})$ such that the sequence $(\langle
d_0,\psi_{2,a}\rangle_a)$ strongly approaches $(\psi_a)$, that is,
$\lim_{a\to 0}||\psi_a-\langle d_0,\psi_{2,a}\rangle_a||_p=0$.}

It is clear that this conjecture is stronger than Conjecture 2.  It
states that conditions $(a)$ and $(b)$ are necessary for a minimizing
sequence. On the other hand, assuming Conjecture 2, they are
sufficient: first note that $\psi_a\equiv\phi_a$ in $L^p(\rR)$
guarantees $\widetilde{\psi}_a\equiv \widetilde{\phi}_a$ in
$L^q(\rR)$, and thus if $||\psi_a||_p$ is normalized to unity and
$||\widetilde{\psi}_a||_q$ has a finite non zero limit, $\lim_{a\to
0}(\cS_q(\psi_a)-\cS_q(\phi_a))=0$. Further, Proposition 1 was proved
assuming $\psi_2(x,y)$ to be independent of $a$. The danger with an
$a$-dependent $\psi_{2,a}$ is that, if $a^2\langle
y^2\rangle_{\psi_{2,a}}$ or $a^{-2}\langle x^2\rangle_{\psi_{2,a}}$ do
not go to 0 for small $a$, the various terms in the series of $\psi_1$
or $\widetilde{\psi}_1$, respectively, overlap and the proposition
does not apply. This danger is avoided by condition $(a)$ since
$\psi_a$ is assumed to approach $d_0$ which consists of well separated
Dirac deltas.

In conclusion, we have presented a set of conjectures on the infimum
and on the minimizers of the functionals $\cS$ and $\cS_q$ in the
space of odd one-dimensional functions. They are based on information
obtained through a simple-minded direct approach, namely, a numerical
minimization. This cannot be made into anything rigorous, since the
numerical procedure might be lead to a relative minimum, rather than
to the absolute one, however this possibility seems quite unlikely to
us since the numerical result has been checked to be stable against
details of the calculation, including changes in the initial
conditions chosen for the minimization.

Although at first sight the numerical result in Figure~1 seems to be
rather irregular, we have hopefully shown in this work that in fact it
is plenty of structure and regularity. The space ${\cal H}(d_0)$ has
proven to have nice properties directly inherited from the map
$d_0\otimes L^p(\rR^2)$ into $L^p(\rR)$. The numerical value of the
infimum of $\cS$ in $\cH$ has been understood as an approximation to
twice the absolute infimum in $L^2(\rR)$ and the numerical minimizer
has been understood as a (singular) double Gaussian
structure. Gaussian functions seem to dominate the entropy
minimization problem both in the whole space and in the odd functions
subspace. Likely, these regularities will open the way for a rigorous
treatment of the problem studied here.


\appendix{}\section{Proof of Lemma 1.}

Let us begin by proving $\psi_1^\prime\equiv\widehat\psi_1^\prime$.
For $p\ge 1$, $||\ ||_p$ is a norm, hence, due to the triangle inequality
\begin{eqnarray}
||\psi_1^\prime-\widehat\psi_1^\prime||_p 
 &\le& b\sum_{\scriptstyle  n,m \atop \scriptstyle m\not=n}
\left(\int_{I_m}\left|\psi_2\left(ax_n,(x-x_n)/a\right) \right|^p\dd
x\right)^{1/p}\,.
\end{eqnarray}
By assumption $\psi_2$ is bounded and fast decreasing at infinity,
thus for any $s$ and $t$ positive, there is a positive $K$ such that
$|\psi_2(x,y)|\le K(1+x^2)^{-s}|y|^{-t}$. Also, for $x\in I_m$,
$|x-x_n|\ge r(|n-m|-\frac{1}{2})$. Therefore, for $s>1/2$ and $t>1$
\begin{eqnarray}
||\psi_1^\prime-\widehat\psi_1^\prime||_p &\le& bKa^t
\sum_n \left(1+a^2x_n^2\right)^{-s}\sum_{m\not=n}
\left(\int_{I_m}|x-x_n|^{-tp} \dd x\right)^{1/p}\, \nonumber \\ &\le&
bKr^{1/p-t}a^t \sum_n\left(1+a^2x_n^2\right)^{-s}\sum_{m\not=n}
\left(|n-m|-\frac{1}{2}\right)^{-t} \,, \nonumber \\ &=& 2bKr^{1/p-t}a^t
\sum_{m\ge 1}\left(m-\frac{1}{2}\right)^{-t} \sum_{n\in\rZ}
\left(1+a^2x_n^2\right)^{-s} \nonumber \\ &\le& K^\prime a^{t-1} \,.
\end{eqnarray}
In the last inequality we have used that the series on $n$ is of order
$a^{-1}$ since $(1+x^2)^{-s}$ is Riemann integrable and the $x_n$ are
equidistantly distributed. The proof of $\psi_1\equiv\widehat{\psi}_1$
is analogous.

Since $\equiv$ is an equivalence relation, it remains only to show
that $\widehat\psi_1\equiv\widehat\psi_1^\prime$.
\begin{eqnarray}
||\widehat\psi_1-\widehat\psi_1^\prime||_p^p &=&
b^p\sum_{n\in\rZ} \int_{I_n}\Big|
\psi_2\left(ax,(x-x_n)/a\right)-
\psi_2\left(ax_n,(x-x_n)/a\right)\Big|^p\dd x \nonumber
\\ &\le& b^pa\sum_{n\in\rZ} \int\Big|
\psi_2\left(ax_n+a^2x,x\right)-\psi_2\left(ax_n,x\right)\Big|^p\dd x\,.
\end{eqnarray}
Again, $\partial\psi_2(y,x)/\partial y$ is a fast decreasing function,
thus, choosing $s>1/2$ and $t>1$, 
\begin{eqnarray}
\left|\psi_2(y+a^2x,x)-\psi_2(y,x)\right| &\le&
\left|\int_y^{y+a^2x}\left|\partial_z\psi_2(z,x)\right|\dd z\right|
\nonumber \\
&\le&
\left|\int_y^{y+a^2x}\frac{K}{(1+x^2)^t(1+z^2)^s}\dd z\right|
\nonumber \\
&\le&
\frac{a^2K|x|}{(1+x^2)^t(1+y^2)^s}\,.
\end{eqnarray}
\begin{eqnarray}
||\widehat\psi_1-\widehat\psi_1^\prime||_p^p &\le&
b^pK^pa^{2p+1}\int\frac{|x|^p}{(1+x^2)^{tp}}\dd x 
\sum_{n\in\rZ}\frac{1}{(1+(ax_n)^2)^{sp}}
\nonumber \\
&\le&
K^\prime a^{2p}
\end{eqnarray}

This completes the proof of the lemma. Note that the conditions
imposed on $\psi_2$ are far more restrictive than actually needed in
the proof.

\begin{figure}
\leavevmode
\hbox{
\hspace{-2cm}
\epsfxsize=14cm
\epsffile{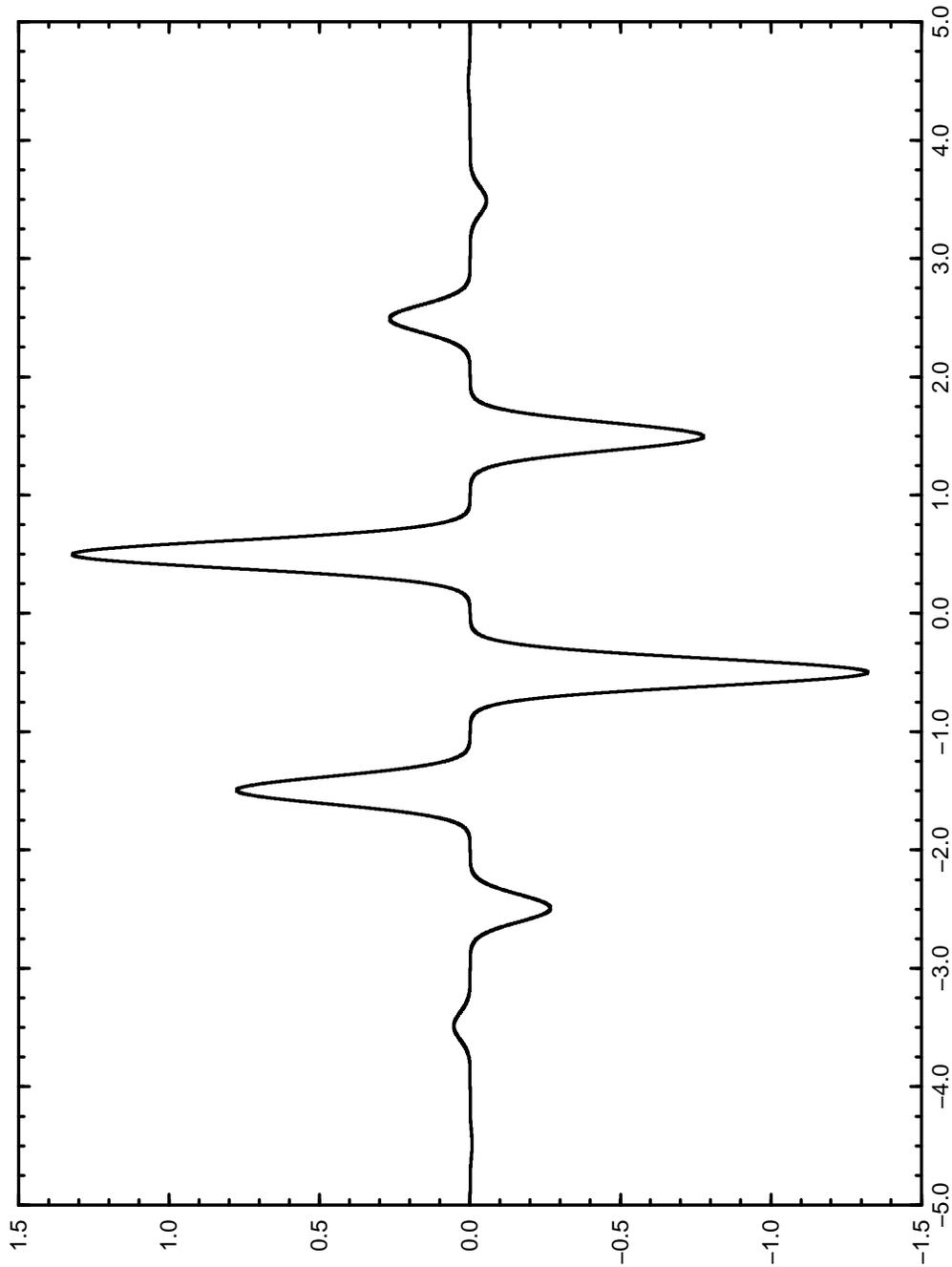}
}
\caption{ Best minimizer of $\cS$ obtained through a 128-dimensional
approximation to $\cH$ (cf. eq.~(\ref{eq:4})). The function is purely
real and also satisfies $\widetilde\psi=+i\psi$. The corresponding
value of $\cS$ is $0.61370581$\,. }
\end{figure}

\end{document}